\documentclass{mem}
\usepackage{natbib}\usepackage{txfonts}\usepackage{balance}
\usepackage{graphicx}
\usepackage[a4paper]{hyperref}
\idline{75}{282}

\begin{document}

\title{
The Mexican Million Models Database: a virtual observatory for gaseous nebulae
}

   \subtitle{}

\author{Morisset, C.}
\offprints{C. Morisset}

\institute{
  Instituto de Astronom\'{\i}a,
  Universidad Nacional Aut\'onoma de M\'exico;
  Apdo. postal 70--264; Ciudad Universitaria;
  M\'exico D.F.; 04510 M\'exico.\\
  \email{Chris.Morisset@gmail.com}
}

\authorrunning{Morisset}

\titlerunning{The Mexican Million Models Database}

\abstract{
The 3MdB (Mexican Million Models database) is a large database of photoionization models for H~II regions. The number of free parameters for the models is close to 15, including the description of the ionizing Spectral Energy Distribution (effective temperature, luminosity, surface gravity, for different type of stellar atmosphere models) and the description of the ionized gas (distance to the ionizing source, density, abundances of the most common elements, dust). The outputs of the models are more than 70 emission line intensities, the ionic fractions and temperatures. All the parameters and outputs are included in the MySQL database, giving the possibility to the user to search into the database for example for all the models that reproduce a given set of observations.

         }
\maketitle{}

\section{Introduction}\label{sec:introduction}

The study of the ionized interstellar medium (in the present case I consider only H~II regions) is mainly based on the analysis of the observed emission line intensities. From line ratios one may determine physical and chemical parameters of the nebulae such as the electron temperature, the electron density and the abundances of the most common elements. The characteristics of the ionizing spectrum (effective temperature, luminosity) can also be derived from the line intensities.

The interaction between the ionizing source and the gas is computed a photoionization code \citep[e.g. Cloudy, see ][]{ferland} allowingto constructu numerical models of H~II regions, including the intensities of the emission lines.
Such models can then be compared to the observations and if all the observables are reproduced one can think that the model is close to a good description of the observed object. One must still be aware that double solution can exist, see Sec.~\ref{sec:non-linearity}.

I present here a new database of photoionization models, which can be used to look for models that are reproducing a given observation or a given catalog of observations. This tool can be understaood as a kind of H~II regions virtual observatory where line intensities from millions of models can be mined.

\section{P-space and O-space}\label{sec:p-space}

\begin{figure*}[t!]
\resizebox{\hsize}{!}{\includegraphics[clip=true]{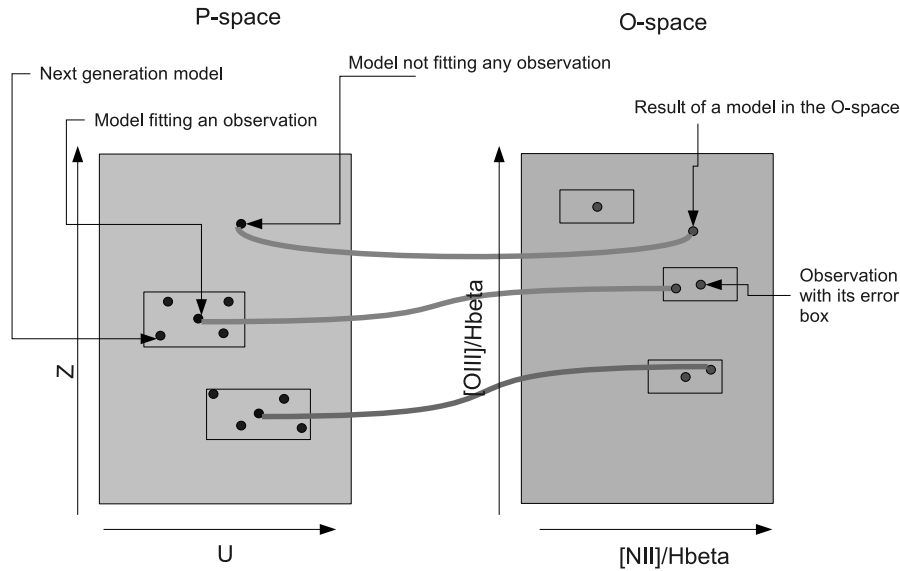}}
\caption{\footnotesize
O-space and P-space in a simple case where the dimensions of both spaces are only 2. The 2 parameters in the P-space are Z and U, the 2 observables in the O-space are the line ratios [NII] and [OIII] over Hbeta.
}
\label{fig:OP-spaces}
\end{figure*}

One can describe a (photoionization) model as a link from the parameter-space (P-space) to the observable-space (O-space). The parameter-space is describing an object in terms of effective temperature, luminosity, size of the nebula, radial density variation, abundances, presence of dust, etc. This can be seen as the set of inputs required to compute the model. The object in the observable-space is described by the set of the emission line intensities. This is also the set of outputs of the photoionization model.

The dimension of the P-space is the number of free parameters needed to describe a model, it can easily reach a value of 15 for 1D models (as when running Cloudy), many more for 3D models where the description of the density distribution is more complexe \citep[using e.g. Cloudy\_3D, see][]{c3d}. The dimension of the O-space is the number of emission line intensities that one can obtain from the photoionization code. It can be seven hundreds of lines! But most of these lines are redundant: their intensities is proportional to another line, e.g. [OIII]4959 and [OIII]5007 or not observed, because of their low signal/noise or because no observation is available in the corresponding wavelength range for a particular object.

In the O-space we find the results of the modeling process (what we classically call the models, projections from the P-space into the O-space using a code) and also the observations of ``real'' objects. Actually, taking into account the error bars around each observed value of emission line intensity transform the observed objects to an hyper-boxes around the observed values (in the O-space).

Fig.~\ref{fig:OP-spaces} illustrats the relation between the P-space and the O-space. The modeling process is represented by the link between the 2 spaces. A model is actually the projection of a set of parameters (a point in the P-space) into a point in the the O-space.

\subsection{Non linearity, degeneracies}\label{sec:non-linearity}
Any point in the P-space transforms into a point into the O-space. The function that transforms a point from P into a point in O is continuous, therefore any shape in the P-space also transforms into a shape in the O-space.
The relation between the shape in the P-space and the corresponding shape in the O-space is far from being linear. 
For example, a rectangule in the P-space does not transform into a rectangular plane in the O-space, but rather into a complex hyper-shape. This is illustrated by Fig.~2 in \citet{Grazy} where a regular grid in the P-space (of 2 dimensions U and Z) transforms into a curved shape into the O-space.

The reverse is also true: a rectangular shape into the O-space is not obtained by a rectangular shape in the P-space: this is why it is not possible to easily obtain the parameters of the models that adjust a given observation (See sec.~\ref{sec:fitting-observations}).

In the case illustrated by Fig.~2 in \citet{Grazy}, the problem is even worst as the projected shape into the O-space of the rectangle from the P-space is an overlapping surface. This leads to a degeneracy, as the same point in the O-space is obtained by 2 different points in the P-space.

\subsection{Fitting an observed object}\label{sec:fitting-observations}

The action of fitting an observation by some models is finding the models which are close to a given observation in the O-space. Considering the errors on the observations, this means finding the models that fall in the hyper-box around the point that represent the object in the O-space. In the case illustrated by Fig.~\ref{fig:OP-spaces}, the fitting models are falling within the rectangle around the observations. Due to the high non-linearity of the transformation between the P- and the O-space, there is no simple way to go from an observation to the set of physical parameters that describe the object.

There are various ways to find the set of values in the P-space that reproduce an observed object (a point in the O-space, or an hypercube if we take the error bars into account):
\begin{itemize}
\item By hand: running models and figuring out what are the effect in O-space of changing something in the P-space.

\item By automatic Khi2 method: for example Cloudy can optimize a set of parameter to fit a set of observations.
\end{itemize}

Generally the two methods above lead to a definition of the ``best'' model fitting the observations of an object.
\begin{itemize}

\item Regular grids of models:
this method can be very useful to see the effects of changing one parameter on the observables.
It gives the possibility of finding various models that fit the same observation (within the errors)
One major problem is that only a few parameters can be changed (5 parameters with 7 values each leads to... 80000 models!)
A second problem is that most of the models are totally useless (in the corners of the hypercube in the P-space, therefore most of the time not corresponding to any observations)

\item Irregular grids of models:
This is the case of a grid that can be adapted to increase the density of models in the P-space in regions where this is useful.
Such an approach needs observations to know which locus in the P-space is ``good'' (it falls in a ``good'' locus in the O-space : where there is observed objects).
For this one can use a kind of genetic algorithm, see next section.
\end{itemize}

\section{A genetic algorithm for the definition of new models}
\label{sec:genetic-algorythm}

To define a genetic algorithm, we must considere two phases: a phase of selection of parents and a phase of reproduction with random evolution, generating children.

The selection of the parent models is performed in the O-space, in the hyper-boxes around the observations, the sizes of the hypercube being the acceptable error on each observable (e.g. emission line intensity). Any model that falls within an hyper-box around an observation is a model selected for the reproduction (it is a parent model).
A new generation of models is generated from the set of parent models. The values of the parameters for the children are determined randomly around the values of the parent models, within a given range. Each parent will generate a given number of children. In the present case, there is no ``sexual'' reproduction in the sense that there is only one parent needed to make children (a kind of unicellular organism reproduction by division and random evolution).
This process is illustrated in Fig.~\ref{fig:OP-spaces}, where new models are represented in the P-space around the parent models, which fit an object in the O-space.
New models in the O-pace can fall around observations that were not fitted before, or be closer to an O-point (leading to a better fit).

The sizes of the different boxes in Fig.~\ref{fig:OP-spaces} play an important role: if the size of the hyper-box in the O-space is small, the number of fitting models is small, but the quality of their fit is good. On the contrary, if the size is big, there will be more models fitting the observations. Some observations that cannot be fitted within a small box can be fitted by models (of smaller quality) with a bigger box. 

On the P-space side, the size of the box is the range in which the parameters will be randomly sorted out. A big P-box will allow an exploration of the P-space, with a possibility of finding models that fit new observations. But given that the new parameters can be quite different from the ``working'' values, the probability of finding better fit is small. On the contrary, defining small P-boxes gives better fits around objects already fitted, leading to a densification of the models around the observed points in the O-space.

\section{The 3MdB}
\label{sec:database}

The Mexican Million Models database is a project of a huge photoionization model database, where the user can search easily and quickly for models that reproduce a given set of observations.

There are more than 15 parameters that can be varied to describe a model: 

\begin{itemize}
\item The ionizing SED can be described as a Planck function (2 parameters: the effective temperature and the luminosity), as a stellar atmosphere model (with various available libraries), in this case the stellar metallicity and the surface gravity may also be provided. There is also a possibility to describe the SED in terms of stellar cluster, with a Starburst99 \citep{Starburst99} ionizing flux (given an age of the burst) or even a description of hundreds of individual stars that form the cluster.
\item The ionized gas: the inner radius of the nebula, the hydrogen density, the abundances of the main elements, the presence of dust (composition, density), a filling factor for the gas.
\end{itemize}

Once the model is computed (using Cloudy) the output files are read and the entry in the database for the model is completed by adding to the parameters the intensities of more than 70 emission lines and all the ionic fractions and temperatures (integrated on the line of sight and on the volume).

An entry in the 3MdB is: a point in the P-space (defined by the values of all the parameters), the corresponding point in the O-space (the values of the observables, i.e. line intensities), plus a set of other characteristics of the models, such as the recombination radius, the ionic fractions and temperatures, the mean ionization parameters, all being parameters that can be useful to the user in understanding the model.

The genetic algorithm described in Sec.\ref{sec:genetic-algorythm} is used to compute the values of the parameters for the new generation models. The observations that are used for the selection of the parent models are from various catalogs, such as metal-poor galaxies from \citet{Izotov}, or the M33 Spitzer observations from \citet{Rubin}.

All the models are in a single table in the database, whatever the set of observations used to select the models: some models computed to fit (optical) SDSS data can be useful for fitting the (IR) M33 HII regions.

The database contains 1,350,000 models (October 2008). The increasing rate of the database is ~350 models/hour.
It presently run on a 2-double-core AMD 64 bits processors computer.

The data are in MySQL tables, driven by IDL routines calling Cloudy, reading the outputs and filling the database.

There is a queuing system with priorities: a set of models can be sent to the queue at any moment, the models with higher priorities being running before the ones with lower priority. This allow the user to quickly run a small grid of models while a larger grid with lower priority is waiting.


\section{The future}\label{sec:future}

\subsection{A User-friendly interface}\label{sec:user-friendly-interf}

The 3MdB will be accessible through a user-friendly interface in a short future. There will be some possibility of selecting the models by any criteria, for example by fitting a given object or set of objects, within observational tolerances.
The actual time needed to search in the whole database for all the models reproducing 10 emission line ratios is only 10 seconds.

\subsection{Virtual Observatory integration}\label{sec:virt-observ-integr}

One of the next evolution of the 3MdB is to insure the interoperability with the emission line databases of HII regions or galaxies. It will be possible to directly search in the 3MdB the list of models that reproduce an object from the VO space.

\begin{acknowledgements}
This project and the participation to the VO congress are partially supported by CONACyT grant 49737.
\end{acknowledgements}

\bibliographystyle{aa}

\end{document}